\documentclass[12pt]{article}
\usepackage{cite}
\usepackage{graphicx}
\usepackage[utf8]{inputenc}
\usepackage[dvips]{epsfig}
\usepackage{graphicx}
\usepackage[T1]{fontenc}
\usepackage{xcolor}
\usepackage{color,soul}
\usepackage{authblk}
\usepackage{amsmath}
\usepackage{float}
\usepackage{hyperref}
\usepackage{bbold}
\date{}
\usepackage{algorithm}
\usepackage{algorithmic}

\date{\today}

\title{\bf Optimal entanglement witness of multipartite systems using support vector machine approach.}
\author[1,2,3]{Mahmoud Mahdian\thanks{mahdian@tabrizu.ac.ir}}
\author[1]{Zahra Mousavi\thanks {z.mosavi1400@ms.tabrizu.ac.ir}}
\date{\today}

\affil[1]{Faculty of Physics, Theoretical and Astrophysics
Department, University of Tabriz, 51665-163 Tabriz, Iran}
\affil[2]{Research Institute for Applied Physics and Astronomy (RIAPA), University of Tabriz, Tabriz, Iran}
\affil[3]{Quantum Technology Center, University of Tabriz, Tabriz, Iran}

\begin{document}
\maketitle
\begin{abstract}

An  entanglement witness (EW) is a Hermitian operator that can distinguish an entangled state from all separable states. We drive and implement a numerical method based on machine learning to create a multipartite EW. Using support vector machine (SVM) algorithm, we construct EW's based on local orthogonal observables in the form of a hyperplane that separates the separable region from the entangled state for two, three  and four qubits in Bell-diagonal mixed states, which can be generalized to multipartite mixed states as GHZ states in systems where all modes have equal size.
 One of the important features of this method is that, when the algorithm succeeds, the EWs are optimal and are completely tangent to the separable region. Also, we generate non-decomposable EWs that can detect positive partial transpose entangled states (PPTES).

\end{abstract}

\section{Introduction}
An important and necessary resource in quantum science and technology is quantum entanglement, which was first described in the works of Einstein \emph{et al}. \cite{Einstein}  and \cite{Schrodinger}  introduced by Schrödinger.
This quantum phenomenon, which challenged the concepts of correlation and caused new interpretations in physics, has
many applications in quantum information and calculations, measurement, cryptography, and teleportation (for an introduction, see e.g.\cite{Nielsen, Bennett,Bennett1,PhysRevLett.67.661}). Therefore, determining whether a quantum state is entangled or not has been one of the most important research fields in recent years\cite{GUHNE20091}, which is a typical classification problem and it has been proven to be a NP-hard, and challenging even for  bipartite systems \cite{Gurvits}.

First, we recall the definitions of entanglement and separability of quantum states.

\emph{\textbf{Definition-1}}: A pure state of an \textit{n}-partite system $|\psi\rangle = |\alpha_1\rangle\otimes|\alpha_2\rangle...\otimes|\alpha_j\rangle$ belonging to the multipartite Hilbert space$(|\psi\rangle \in \mathcal{H}= \mathcal{H}_{d_1}\otimes\mathcal{H}_{d_2}...\otimes\mathcal{H}_{d_n} $is called $j$-separable $( j \in 2, 3, \ldots,\textit{n} \    s.t. \  d_1 d_2 \ldots d_j =\textit{n} ) $ if and only if it can be written as a product of $j$ subsystem states in the
following form:
\begin{equation}\label{eq:state11}
\rho_{sep}^j = |\alpha_1\rangle\langle\alpha_1|\otimes|\alpha_2\rangle\langle\alpha_2|...\otimes|\alpha_j\rangle\langle\alpha_j|,
\end{equation}
where $|\alpha_i\rangle \in \mathcal{H}_{d_i}$ and $d_i$'s are dimensions of Hilbert space and it is \textit{fully separable} if \textit{j} = \textit{n}. A
multipartite quantum mixed state, which is a bounded operator acting on the Hilbert Space $\mathcal{H}$, is fully separable if its density matrix is a convex sum of pure product states \cite{Horodecki22,Gabriel}, such as
\begin{equation}\label{eq:state2}
\rho = \sum _i p_i \varrho_i= \sum _i p_i \varrho_i^{(1)}\otimes\varrho_i^{(2)}...\otimes\varrho_i^{(n)}  , \\ \sum_i p_i =1 , \\ p_i\geq 0 \textcolor{red}{,}
\end{equation}
otherwise, $\rho$  is called entangled. In our calculations, we will consider the multipartite systems as N-qubit space which  is N-fold tensor
product of $ (C^2) ^{\otimes N}$.\\
Most applications of quantum computing and information are based on the non-local nature of quantum mechanics and entanglement. Therefore, determining whether a state is entangled is very important for quantum information applications. In low-dimensional Hilbert spaces, a necessary and sufficient condition  exists for separability. In the Hilbert space with $\mathcal{H}_2\otimes\mathcal{H}_2$ and $\mathcal{H}_2\otimes\mathcal{H}_3$ dimensions, there are conditions for the separability of states called the positive partial transpose (PPT) criterion or Peres-Horodecki criterion \cite{Peres,Horodecki}. But for quantum systems in higher dimensions, PPT's condition is necessary, but not sufficient, this means there can be quantum states where the partial transpose is positive, but the states are still entangled (PPT entangled states, also called \textit{bound entangled states}.) These states are very difficult to detect \cite{Horodecki12}.

Another method for detecting the entanglement of these states in high dimensions is the use of EW's which are operators that can detect entangled states. Optimal EW operators are able to
detect \textit{all} entangled states, including the PPT-entangled states, also called bound entangled states.

\emph{\textbf{Definition-2}}:
An EW \textit{W} acting on Hilbert space $\mathcal{H}$ is a Hermitian operator ($W=W^\dagger$), such that (i) Tr$(W \rho_{sep}) \geq 0 $ for every separable state $\rho_{sep}$ in Banach space (Hilbert space of bounded operators),  and (ii) there exists an entangled state $\rho_{ent},$ Tr$(W \rho_{ent}) < 0 $ or \textit{W} has at least one negative eigenvalue \cite{Lewenstein,Jafarizadeh,PhysRevA.78.032313}. \\
If the expected value of the EW on the density matrix is negative, then it is an entangled state and $W$ is an EW which can detect this state. Having an entangled witness for each entangled state is a result of Hahn-Banach theorem and that the subspace of separable states is convex and closed. From the geometrical point of view, the EWs are a hyperplane that separate the separable states from the entangled
state, and we can represent this hyperplane in a two-dimensional picture as a line locally tangent to
the set of separable states.

\emph{\textbf{Definition-3}}:
We call an EW \textit{W optimal} if and only if it is impossible to construct a \textit{new} EW of the form $ W_{new}=(1+\epsilon)W-\epsilon \mathcal{T}$ for some
positive operator $\mathcal{T}$  and scalar $\epsilon > 0$; that is, if no such $\mathcal{T}$ and $\epsilon$  can be found to create a valid new
EW this way, then the original \textit{W} is an optimal EW. Since every EW must have a nonnegative expectation value on
the separable region, the positive operator $\mathcal{T}$ exists on the condition that it is orthogonal to the kernel
of the witness operator \cite{lewenstein2000optimization}.

\emph{\textbf{Definition-4}}:
Based on the partial transpose on the subspaces of the multipartite systems, the EW's can be divided into two different classes, which are: the class of decomposable EW's and the class of non-decomposable EW's. An EW is decomposable if there are positive operators $\mathcal{K}$ and $\mathcal{O}$ that the EW can be expanded based on them.
\begin{equation}\label{decom1}
W=\mathcal{K}+\sum_{i\subset N} \mathcal{O}_i^{T_i},
\end{equation}
where N=$\{1,2,3...n\}$ and $T_i$ indicates partial transposition with respect to subsystems and the EW is nondecomposable if it cannot be written in form \ref{decom1}.

In recent years, the intersection between machine learning and quantum computing has attracted a lot of attention \cite{Arunachalam,Ciliberto}. This field can be considered from two aspects: firstly, using quantum computing's fast computation and data storage capabilities to improve classical machine learning performance \cite{Chrisley,Siheon}, and Secondly, using machine learning algorithms in quantum information and computing \cite{Torlai,PhysRevX.8.031086,Ming,Carleo}. The detection of entanglement using machine learning has recently received attention due to its high accuracy and efficiency for high-dimensional Hilbert spaces \cite{Naema,PhysRevResearch.3.033135,Harney_2021,PhysRevA.98.012315,8764462}.

This paper investigates the classification of quantum states into entangled and separable states using machine learning techniques. We obtain training data by performing non-local measurements on quantum systems. An SVM-based separability-entanglement classifier is then used to construct an entanglement witness EW that can detect entangled states. In this supervised learning method, we feed the classifier with a large number of measured data as well as the corresponding class labels (entangled or separable) and finally the classifier to predict the class labels of new quantum states. In this work, we obtain EW's for two, three and four-qubit systems. One of the advantages of this method is that the EW's are optimal, and we also obtain EW's that can detect PPTES which can be generalized to multipartite states.\\
The paper is organized as follows. In Section 2, we introduce our SVM model and separability-entanglement classifier. In Section 3, we construct optimal EW's for two, three and four-qubit quantum states with the SVM method. Finally, Section 4 presents the conclusions of the study.

\section{SVM}

One of the most powerful supervised learning algorithms is SVM. It is used for various tasks including classification and regression. In this algorithm, data classification can be done by obtaining an optimal hyperplane with the largest margin. This approach is also applicable to multi-class classification problems. In the data classification algorithm, the machine learns to determine the most probable class of an unknown data $\tilde{x} \in C^N$ according to the training data D=$\{(x_i,y_i) | x_i \in C^N , y_i \in (0,1,...)\}$ where 
\( x_i \) represents the \( i \)-th feature vector in \( C^N \), which is a vector of \( N \) complex numbers. \( y_i \) is the class label for \( x_i \), an integer in \( \{0, 1, \ldots, K-1\} \), where \( K \) is the total number of classes.
\cite{support,svm}.
In the case that the training data has some error, such that some of them are not in their class, then we have SVM with soft margin as:

\begin{equation}
\begin{array}{rrclcl}
\displaystyle \min_{w,b,\xi} & \multicolumn{3}{l}{\frac{1}{2}w^{t}w+C\sum_{i=1}^{N}{\xi_{i}}},\\
\textrm{s.t.} & y_{i}(w x_{i}+b)\geq 1-\xi_{i}, \forall i,\\
&\xi_i\geq0  , i=(1,...,m).  \\
\end{array}
\end{equation}

\( w \) is weight vector that defines the hyperplane and \( b \) is bias term that shifts the hyperplane. Penalty term \( C \sum_{i=1}^{N} \xi_i \), $\xi_i$ as the slack variables for misclassification, where \( C > 0 \) is a regularization parameter that balances the trade-off between maximizing the margin and minimizing misclassification.

Since we are dealing with separable and entangled states, this is a two-class problem, i.e., $y_i \in (1,-1)$, that we label separable states with $y=1$ and entangled states with $y=-1$. This binary classification can be extended to multi-class problems using techniques like one-versus-all or one-versus-one with multiple binary SVMs\cite{Giuntini}. According to the correspondence between SVM and the EW, we obtain a hyperplane that can separate the separable states,i.e., Tr$(W \rho_{sep}) \geq 0$ from the entangled states, i.e., Tr $(W \rho_{ent}) < 0$.

In the N-qubit Hillbert space, we consider the Hermitian operator of EW as follows, in which the measurement is applied locally on the subsystems.

\begin{equation}\label{eq:state44}
W = \sum_{i_1,i_2,...,i_N} c_{i_1,i_2,...i_N} \sigma_{i_1} \otimes \sigma_{i_1}...\otimes \sigma_{i_N} ,
\end{equation}
where $\sigma_{i_j}$ are Pauli matrices \{$\sigma_x,\sigma_y, \sigma_z$\ and also identity matrice $I$\}. First, in order to train data related to the class, we generate separable and entangled states, in which quantum circuits can be used. Then, we measure the Pauli operators on each of these states and record the measurement results as features to train SVM:
\begin{equation}\label{p111}
x_{\vec{i}}= \text{Tr}(  \sigma_{i_1} \otimes \sigma_{i_2}...\otimes \sigma_{i_N} \rho)=\langle  \sigma_{i_1} \otimes \sigma_{i_2}...\otimes \sigma_{i_N}\rangle,
\end{equation}
 and expectation values of EW is
 \begin{equation}\label{p2}
\langle W\rangle= \sum_{i_1,i_2,...,i_N} c_{i_1,i_2,...i_N} x_{\vec{i}}=\sum_{i_1,i_2,...,i_N} c_{i_1,i_2,...i_N}\langle  \sigma_{i_1} \otimes \sigma_{i_2}...\otimes \sigma_{i_N}\rangle,
\end{equation}
where $\vec{i}=(i_1,i_2,...i_N)$.

By performing optimization in SVM, we calculate the values of coefficients $c_{i_1,i_2,...i_N}$ and having these coefficients according to equation \ref{eq:state44}, we can obtain the optimal hyperplane for the classification of entangled and separable states. We define the hinge loss function for our soft margin SVM by

\begin{equation}\label{p33}
\mathcal{L}= \max (0,1- y_i (w^T x_i-b))=\max (0,1- y_i (\sum_{\vec{i}} c_{\vec{i}} x_{\vec{i}}-b)),
\end{equation}
where \( b \) and  \( w \) are bias term and weight vector in the decision function, respectively.
 with following classifier:

\begin{equation}
\begin{array}{rrclcl}
\displaystyle y= \sum_{\vec{i}} c_{\vec{i}} x_{\vec{i}},\\
\textrm{s.t.} &y \geq0 \ \ if \  \ seperable, \\
&y< 0 \ \ if \  \ entangled.  \\
\end{array}
\end{equation}

\begin{figure}
	\centering
	\includegraphics[width=0.7\linewidth]{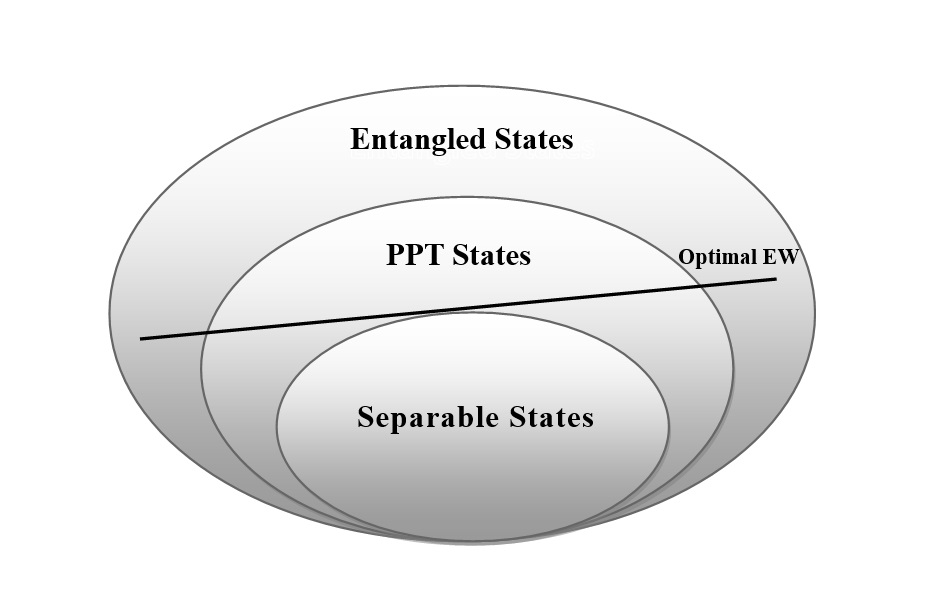}
	\caption{The figure illustrates the convex separable states, PPT entangled states, and entangled states. Based on this geometric structure, a hyperplane is defined as the entanglement witness (EW). This hyperplane separates the separable states from the region containing both PPT entangled states and other entangled states.}
	\label{fig:screenshot-2024-04-11-162402}
\end{figure}

\section{APPLICATIONS}

In this section, we obtained EW's using SVM for two, three, and four-qubit.
These EWs can be generalized to multipartite systems, and we will also prove their optimality.

\subsection{Two-qubit EW}

A class of bipartite quantum states (two-qubit acting on $ C^2 \otimes C^2$) that are widely used in quantum computing and information are Bell-diagonal states which are in the following density matrix:

\begin{equation}\label{eq:state1}
\rho = \sum_{ij} \lambda_{ij} |\varphi_{ij}\rangle \langle \varphi_{ij}|=\frac{1}{4}( I+\sum_{i=1}^3 c_i \sigma_i \otimes \sigma_i) .
\end{equation}

The four Bell-states are

\begin{equation}\label{bel1}
|\varphi_{ij}\rangle=\frac{1}{\sqrt{2}}(|0,j\rangle+(-1)^i |1,1\oplus j\rangle) \ where \ i,j \in \{0,1\},
\end{equation}
with eigenvalues of $\lambda_{ij}=\frac{1}{4}(1+(-1)^i c_1-(-1)^{i+j} c_2+(-1)^j c_3)$ and the $\sigma_i$'s are Pauli operators. From the geometric point of view, the density matrix \ref{eq:state1} can be considered as a tetrahedron, where the pure entangled states are placed as  extremal points on its four corners. Bell's mixed state, which can be considered as a convex combination of pure states, is placed inside this tetrahedron. Also, the separable states that are in the form of a convex set are also placed in the form of an octahedron inside the entangled region. Here, we focus on Werner states a special class of quantum states formed by the convex combination of Bell-diagonal states and the maximally mixed state. Utilizing the SVM method, we analyze two categories of Werner states, entangled and separable, to derive optimal EW for entanglement detection. 
One of the advantages of using the machine learning technique and especially SVM in this paper is that all EWs are completely tangent to the separable region and are optimal, and we will prove this further.
In this paper, we consider the Werner states which are invariant under the unitary operation of $U\otimes U$. These states are very interesting because they are a combination of a maximum entangled state and a fully mixed state, which is actually a model of white noise that is used to determine the robustness of the entanglement.
Now, we introduce the state where we created the training data. The Werner state with the free parameter p is given by:
\begin{equation}\label{mat12}
\rho^W_{Bell(i,j)}=p |\varphi_{ij}\rangle \langle \varphi_{ij}| + \frac{1-p}{4} \mathbb{1}_4 ,  0\leq p \leq 1,
\end{equation}

when $p > \frac{1}{3}$, it is entangled state and $|\varphi_{ij}\rangle$ can be one of the bell states \ref{bel1}. $\mathbb{1}_4$  is the unit matrix in the two-qubit space.

\begin{figure}
	\centering
	\includegraphics[width=0.8\linewidth]{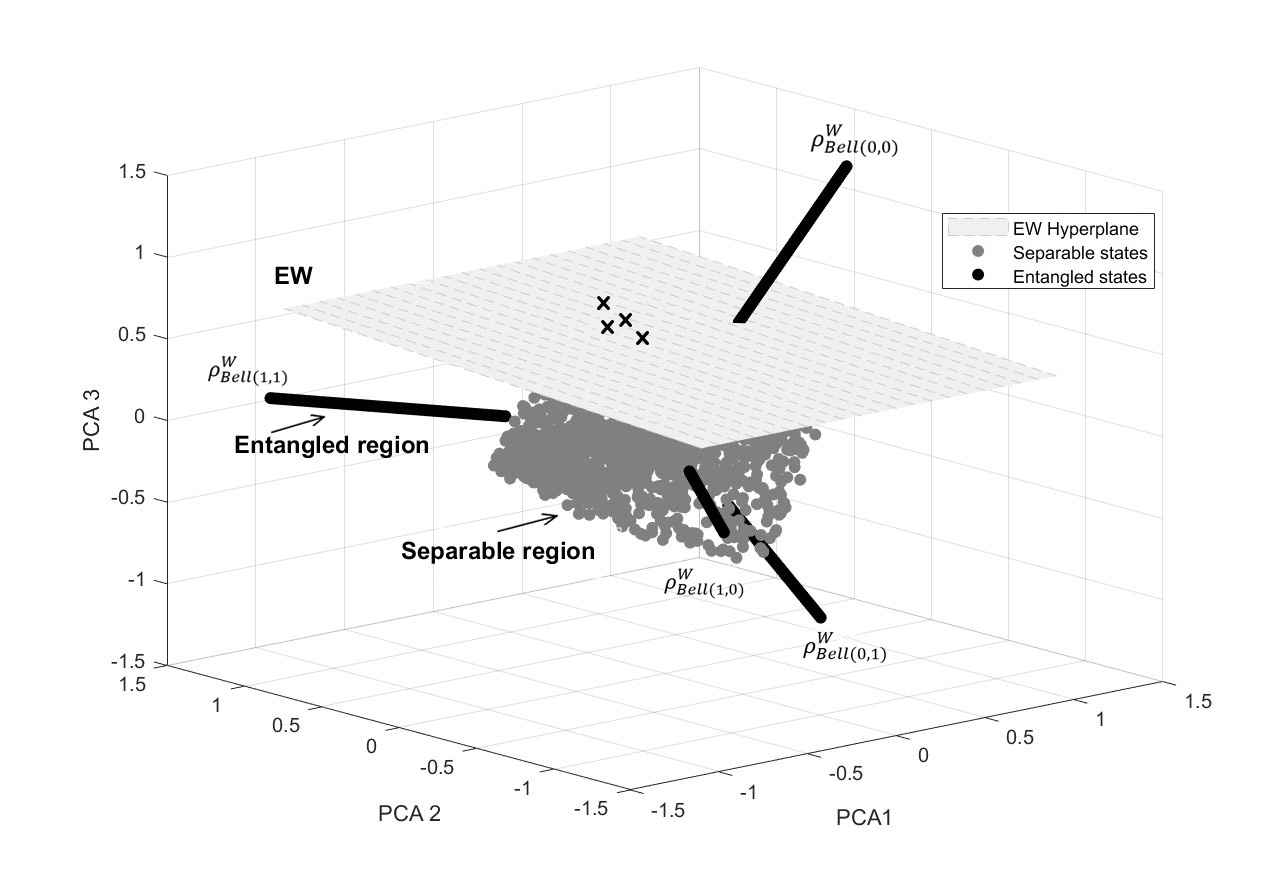}
	\caption{Principal component analysis (PCA) plots of three principal components of separable and entangled states of Bell-diagonal state. PC1, principal 1, PC2, principal 2, PC3, Principal 3 and Optimal entangled witness. The hyperplane in the figure is defined to be perpendicular to the specified points, providing a clear separation between the two types of states in the PCA space.
    For the Bell state $\rho_{Bell(0,0)}^{W}$, we calculate the corresponding EW. The "x" markers represent the tangent states to the EW hyperplane, providing a visual representation of states that lie on the boundary between separable and entangled regions.}
	\label{fig:2qubit}
\end{figure}

we calculate a witness for two-qubit and for separable states that has zero value mentions with "x" markes in fig.

Using the Werner state density matrix \ref{mat11}, we constructed the training data and the EW's for four different regions where the maximum Bell entangled states are located at their vertices. For one of Bell states , the EW is as follows:

\begin{equation}
	\resizebox{\textwidth}{!}{
		$\begin{aligned}
			W_{1} = \begin{bmatrix}
				(0.0545) & (-0.0019+0.0017  \textbf{i}) & (-0.0098-0.0014\textbf{i}) & (-0.5066+0.0053\textbf{i}) \\
				(-0.0019-0.0017\textbf{i}) & (0.4134) & (-0.0091+0.0195\textbf{i}) & (0.0199-0.0059\textbf{i}) \\
				(-0.0098+0.0014\textbf{i}) & (-0.0091-0.0195\textbf{i}) & (0.4793) & (-0.0052+0.0076\textbf{i}) \\
				(-0.5066-0.0053\textbf{i}) & (0.0199+0.0059\textbf{i}) & (-0.0052-0.0076\textbf{i}) & (0.0526) \end{bmatrix}.
		\end{aligned}
		$
	}\label{wit22}
\end{equation}

Where Tr$(W_1 |\varphi_{00}\rangle \langle \varphi_{00}| )< 0 $. As in Figure 44, it can be seen that the EW is made using maximum Bell state $ |\varphi_{00}\rangle=(|0,0\rangle+|1,1\rangle)$ and therefore, because EW is optimal, it recognizes all the entangled states according to matrix \ref{mat11} for all p's.
We know PCA is indeed a powerful dimensionality reduction technique, but its applicability in quantum information must be carefully evaluated, especially  where all singular values hold significant importance. We used SVM, to construct EWs. Instead of relying solely on PCA, which assumes that certain dimensions can be disregarded without losing critical information, we demonstrates that SVM directly address the classification of quantum states into separable and entangled categories and our method construct optimal EWs by considering the full structure of the data, ensuring that no critical information is lost.
Moreover, the application of PCA in visualizing separable and entangled states, is a valuable preliminary step. However, here we underscores the necessity of rigorous testing across diverse quantum systems, including high-dimensional spaces and  where partial data reduction is not feasible. By leveraging machine learning techniques alongside PCA, the robustness and accuracy of quantum state analysis are enhanced. This approach, as detailed in the paper, provides a strong foundation for addressing the limitations of PCA and showcases how advanced computational methods can complement each other to tackle the intricacies of quantum information science effectively.
We drew feature vectors in 3D space according to Figure \ref{fig:2qubit}. The figures drawn based on the first, second and third components of PCA, which are (PC1) has the highest variance and the second principal component (PC2) has a second highest variance and so on.

To prove optimality of EW, we consider a general two-qubit pure product state as :

\begin{eqnarray}
|\nu\rangle= \bigotimes _{i=1}^2\left(
\begin{array}{cccc}
\cos\frac{\theta_i}{2} \\
e^{i\alpha_i}\sin\frac{\theta_i}{2} \\
\end{array}
\right).
\end{eqnarray}
We have previously detailed the witness operator form as presented in the equation ~\ref{p22} and
 for two-qubit is:

  \begin{equation}\label{p22}
 W = \sum_{i_1,i_2} c_{i_1,i_2} x_{\vec{i}}=\sum_{i_1,i_2} c_{i_1,i_2}  \sigma_{i_1} \otimes \sigma_{i_2},
 \end{equation}
 and 
 \begin{equation}\label{p222}
 	\text{Tr} (W |\nu\rangle \langle \nu |) =	\text{Tr} ( \sum_{i_1,i_2} c_{i_1,i_2}  \sigma_{i_1} \otimes \sigma_{i_2}|\nu\rangle \langle \nu |).
 \end{equation}

By selecting the following angles in Tab.\ref{tab:BELL 1 OPTIMIZATION TABLE}, we find that \( \text{Tr}(W_1 |\nu\rangle \langle \nu|) = 0 \) for $\rho_{Bell(0,0)}^{W}$. These states are denoted by "x" marks in Fig.\ref{fig:2qubit}.

\begin{align*} 
    & \text{Tr}(W_1 |\nu\rangle \langle \nu |) =\frac{1}{4} \Bigg( 
    2 \cos^2\left(\frac{\theta_2}{2}\right) 
    \Big[
        e^{i \alpha_1} 
        \Big(
            (-0.0019 - 0.0106 i) (-1 + \cos(\theta_1)) 
            + (0.027 + 0.0074 i) \sin(\theta_1)
        \Big) \\
        &\quad+ (0.009 - 0.0226 i) \sin(\theta_1)
    \Big] \\
    &\quad+ 2 \cos^2\left(\frac{\theta_1}{2}\right) 
    \Big[
        (-0.9061 ) (1 + \cos(\theta_2)) 
        + e^{i \alpha_2} 
        \Big(
            (-0.0019 + 0.0106 i) (-1 + \cos(\theta_2)) \\
            &\qquad+ (0.009 + 0.0226 i) \sin(\theta_2)
        \Big) 
        + (0.027 - 0.0074 i) \sin(\theta_2)
    \Big] \\
    &\quad+ \sin(\theta_1) 
    \Big[
        e^{i \alpha_2} 
        \Big(
            (0.0156 + 0.0004 i) (-1 + \cos(\theta_2)) 
            + (0.9317 ) \sin(\theta_2)
        \Big) \\
        &\qquad+ \Big(
            (-0.0659 - 0.0182 i) 
            - (0.0091 + 0.0195 i) (-i \cos(\alpha_1) + \sin(\alpha_1))
        \Big) \sin(\theta_2)
    \Big] \\
    &\quad+ e^{i \alpha_1} 
    \Big[
        \Big(
            (0.028 + 0.0088 i) (-1 + \cos(\theta_1)) 
            + (0.8732 - 0.0091 i) \sin(\theta_1)
        \Big) \sin(\theta_2) \\
        &\qquad- e^{i \alpha_2} 
        \Big(
            2 
            \Big[
                (1.1203 ) (-1 + \cos(\theta_1)) 
                + (0.028 - 0.0088 i) \sin(\theta_1)
            \Big] \sin^2\left(\frac{\theta_2}{2}\right) \\
            &\qquad+ \Big[
                (-0.0156 + 0.0004 i) (-1 + \cos(\theta_1)) 
                + (0.0659 - 0.0182 i) \sin(\theta_1)
            \Big] \sin(\theta_2)
        \Big)
    \Big]
    \Bigg).
    \label{wittg}
\end{align*}

\begin{table}[H]
	\centering
	\begin{tabular}{|c|c|c|c|c|}
		\hline
		$\theta_1$ & $\theta_2$ & $\alpha_1$ & $\alpha_2$ & Tr$ (W_1 |\nu\rangle \langle \nu |)$ \\
		\hline
		$1.9840$&  $1.9840$&  $2.9760$&  $3.3070$& $0$ \\
		\hline
		$1.8190$ & $1.6530$& $3.9680$&$2.6460$& $0$ \\
		\hline
		$1.6530$& $1.9840$& $3.9680$& $1.9840$& $0$ \\
		\hline
		$1.8190$&$1.6530$& $3.9680$& $2.6460$& $0$ \\
		\hline
	\end{tabular}
	\caption{We identify specific states defined by 
         angles $\theta$ and $\alpha$ are tangent to the EW hyperplane.}
	\label{tab:BELL 1 OPTIMIZATION TABLE}
\end{table}

and we get the following product pure states:

\resizebox{\textwidth}{!}{
	\begin{minipage}{\textwidth}
		\begin{align*}
			|\nu_1\rangle &= (0.2992|00\rangle + (-0.4516 - 0.0753\textbf{i})|01\rangle + (-0.4516 + 0.0754\textbf{i})|10\rangle + (0.7007 - 0.0001\textbf{i})|11\rangle), \\
			|\nu_2\rangle &= (0.6606|00\rangle + (-0.0422 + 0.5126\textbf{i})|01\rangle + (-0.0355 - 0.4299\textbf{i})|10\rangle + (0.3358 - 6.2241 \times 10^{-5}\textbf{i})|11\rangle), \\
			|\nu_3\rangle &= (0.6606|00\rangle + (0.3405 - 0.2649\textbf{i})|01\rangle + (0.4060 + 0.3157\textbf{i})|10\rangle + (0.3358 - 6.2241 \times 10^{-5}\textbf{i})|11\rangle),\\
			|\nu_4\rangle &= (0.3705|00\rangle+(-0.2277 + 0.5193\textbf{i})|01\rangle+(	-0.2726 - 0.2959\textbf{i} )|10\rangle+(0.5822 - 0.2002\textbf{i})|11\rangle).		
		\end{align*}
	\end{minipage}
}\\ \\

Also, a positive operator can be considered as a linear combination of projection operators with the positive coefficients, $\mathcal{P}=\sum_i \lambda_i P_i$.
If we assume  $P_i=|\omega\rangle \langle\omega|$ that $ |\omega\rangle=\sum_{ij=\{0,1\}} \beta_{ij}|i,j\rangle$ , for the pure states obtained by using the relation Tr$(W |\nu\rangle \langle\nu|) =0 $, then, all the coefficients of  $\beta_{ij}$  are zero

\begin{equation}\label{eq:state}
\langle \nu_1 | \omega\rangle=0, \langle\nu_2 | \omega\rangle=0,\langle\nu_3 | \omega\rangle=0,\langle\nu_4 | \omega\rangle=0,
\end{equation}

and satify Tr$(\mathcal{P} |\nu_i\rangle \langle \nu_i|)=0$, which means that there is no positive operator in this case, so our EW is an optimal witness.

\begin{figure}[H]
	\centering
	\includegraphics[width=0.8\linewidth]{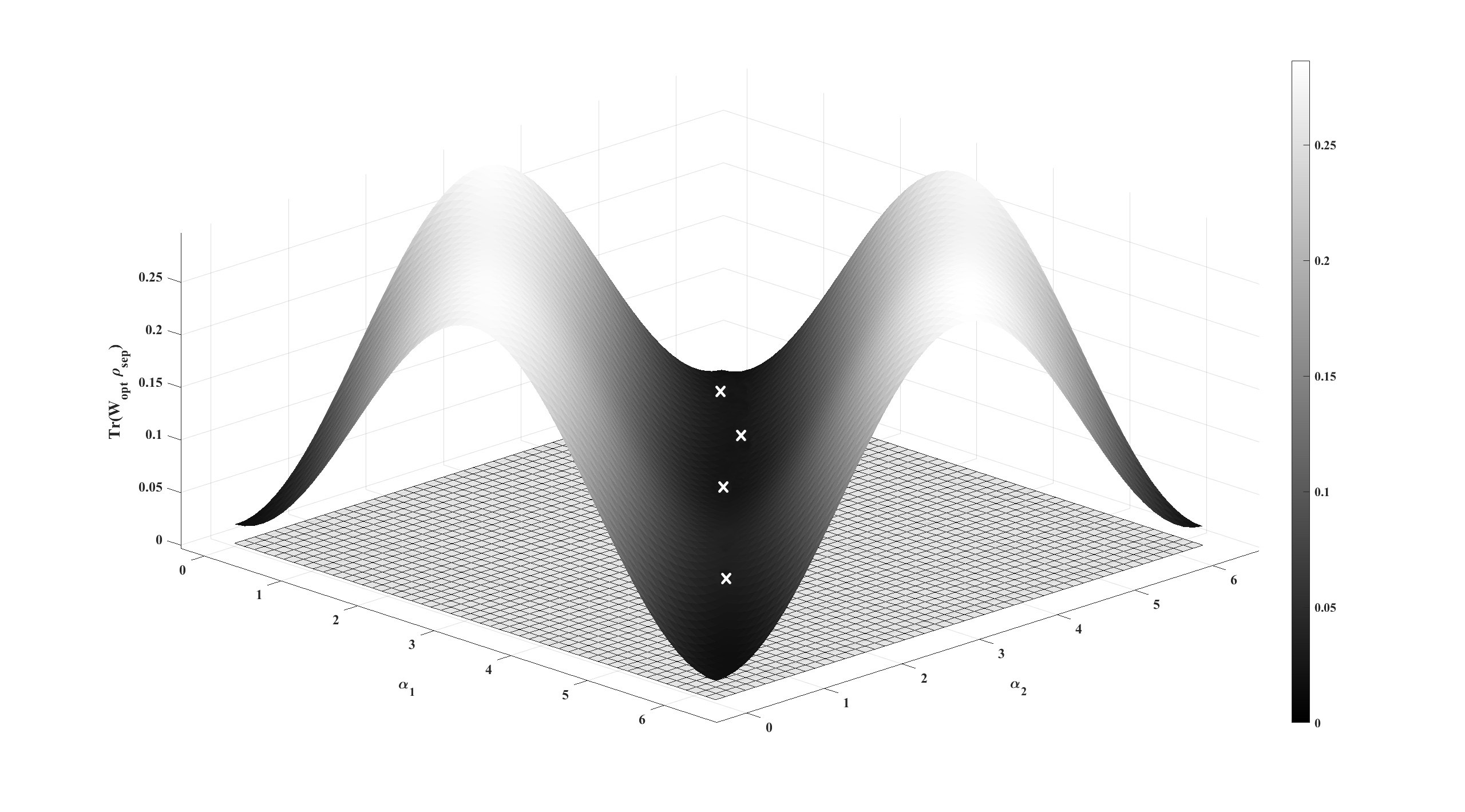}
	\caption{Separable states for different values of $\alpha$ and $\theta$. The separable states that are tangent to the hyperplane or EW are shown in the figure. The introduced witness is tangent with zero value of Tr$(W_{opt} \rho_{sep})$  to the specified points in the figure.}
	\label{fig:optimal-black}
\end{figure}

\subsection{Three-qubit EW's}

Unlike the two-qubit state, where the quantum states are entangled and separable states that can be classifid by measurement, three-qubit systems have a more complex structure and these states can be classified as fully seperable, biseprable, and full entangled called Greenberger-Horne-Zeilinger (GHZ) and W states, i.e., $|GHZ\rangle=\frac{1}{\sqrt{2}}(|000\rangle+|111\rangle)$ and $|W\rangle=\frac{1}{\sqrt{3}}(|001\rangle+|010\rangle+|100\rangle)$. These two entangled states are not transformed into each other using  Local Operations and Classical Communication (LOCC)\cite{Lewenstein11,Vidal}.\\
One of the widely used criteria for detecting entanglement is PPT criterion , which is a necessary and sufficient condition for low dimensions such as two-qubit space and qubit in qutrit spaces. For the higher dimensions, condition PPT is the only necessary condition, and for these states, there may be entangled states that have positive partial transpose. So, the PPT criterion is not sufficient for separability, and for these states, EW can be used to detect PPTES.\\
The Werner density matrix for three-qubit is as follows:
\begin{equation}\label{mat11}
\rho_{GHZ(i,j,k)}^{W}=p |\psi_{ijk}\rangle \langle \psi_{ijk}| + \frac{1-p}{8} \mathbb{1}_8 ,  0.6\leq p \leq 1,
\end{equation}
where,$\psi_{ijk}$ is one of the three-qubit Bell states. Here we obtain EW for state $|\psi_{000}\rangle=\frac{1}{\sqrt{2}} (|000\rangle+|111\rangle)$ as follows using SVM:

\begin{equation} \label{w3}
	\resizebox{\textwidth}{!}{
		$\begin{aligned}
			\mathbf{W}_{GHZ} = \begin{bmatrix}
				(0.053) & (-0.005-0.008\textbf{i}) & (-0.001+0.004\textbf{i}) & (-0.013+0.012\textbf{i}) & (0-0.008\textbf{i}) & (-0.011-0.006\textbf{i}) & (-0.002-0.018\textbf{i}) & (-0.373-0.004\textbf{i}) \\
				(-0.005+0.008\textbf{i}) & (0.162) & (0.005-0\textbf{i}) & (-0.008-0.010\textbf{i}) & (0.016+0.013\textbf{i}) & (-0.008-0.009\textbf{i}) & (0.011+0.004\textbf{i}) & (0.022-0.004\textbf{i}) \\
				(-0.001-0.004\textbf{i}) & (0.005+0.001\textbf{i}) & (0.164) & (-0.012+0.005\textbf{i}) & (-0.008+0.003\textbf{i}) & (0.009-0j) & (0.009-0.008\textbf{i}) & (0.009+0.001\textbf{i}) \\
				(-0.013-0.012\textbf{i}) & (-0.008+0.010\textbf{i}) & (-0.012-0.00\textbf{i}) & (0.126) & (-0.005+0.004\textbf{i}) & (-0.006-0.007\textbf{i}) & (0.005+0.001\textbf{i}) & (0.005+0.001\textbf{i}) \\
				(0+0.008\textbf{i}) & (0.016-0.013\textbf{i}) & (-0.008-0.003\textbf{i}) & (-0.005-0.004\textbf{i}) & (0.159) & (0.007-0.008\textbf{i}) & (-0.006-0.008\textbf{i}) & (0.011-0.007\textbf{i}) \\
				(-0.011+0.006\textbf{i}) & (-0.008+0.009\textbf{i}) & (0.009+0\textbf{i}) & (-0.006+0.007\textbf{i}) & (0.007+0.008\textbf{i}) & (0.135) & (0.010+0.012\textbf{i}) & (-0.001+0\textbf{i}) \\
				(-0.002+0.018\textbf{i}) & (0.011-0.004\textbf{i}) & (0.009+0.008\textbf{i}) & (0.005-0.001\textbf{i}) & (-0.006+0.008\textbf{i}) & (0.010-0.012\textbf{i}) & (0.147) & (-0.001-0.008\textbf{i}) \\
				(-0.373+0.004\textbf{i}) & (0.022+0.004\textbf{i}) & (0.009-0.001\textbf{i}) & (0.005-0.001\textbf{i}) & (0.011+0.007\textbf{i}) & (-0.001-0\textbf{i}) & (-0.001+0.008\textbf{i}) & (0.051) \end{bmatrix}
		\end{aligned}
		$
	},
\end{equation}
To prove the optimality of EW, we consider the following three-qubit product states

\begin{eqnarray}
|\nu\rangle= \bigotimes _{i=1}^3\left(
\begin{array}{cccc}
\cos\frac{\theta_i}{2} \\
e^{i\alpha_i}\sin\frac{\theta_i}{2} \\
\end{array}
\right).
\end{eqnarray}

After some calculations, The following table shows the product states for the values of different angles, which are located on the hyperplane and belong to $Ker(W)$. We apply the same method used for two-qubit systems to explore three-qubit and four-qubit systems.

\begin{figure}[H]
	\centering
	\includegraphics[width=0.9\linewidth]{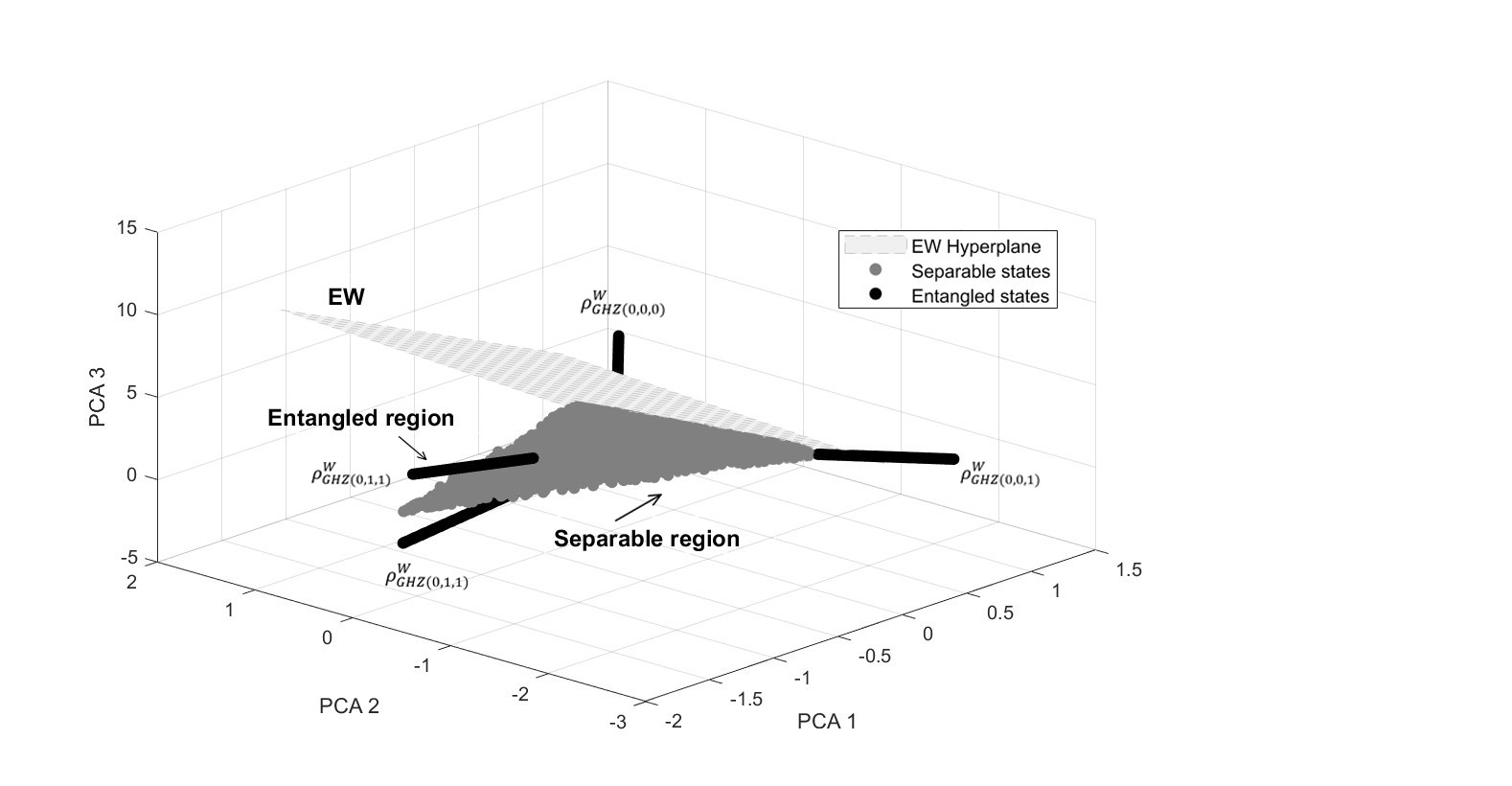}
	\caption{Principal component analysis (PCA) plots of three principal components of separable and entangled states of three-qubit states. PC1, principal 1, PC2, principal 2, PC3, Principal 3 and Optimal entangled witness. The hyperplane is derived using the SVM algorithm for a four-qubit density matrix $\rho_{GHZ(0,0,0)}^{W}$. The calculated EW corresponding to this density matrix is presented in EW \ref{w3} }
	\label{fig:3qubit}
\end{figure}

\begin{table}[H]
	\centering
	\begin{tabular}{|c|c|c|c|c|c|c|}
		\hline
		$\theta_1$ & $\theta_2$ &$\theta_3$ & $\alpha_1$ & $\alpha_2$ & $\alpha_3$& Tr$(W_{GHZ} |\nu\rangle \langle \nu |)$ \\
		\hline
		$1.57$& $1.53$& $1.64$& $4.91$& $5.26$& $2.38$&$0$ \\
		\hline
		$1.57$& $1.53$& $1.64$&$4.91$ &$5.26$ & $2.39$&$0$ \\
		\hline
		$1.57$ & $1.52$& $1.64$ & $4.92$ &$5.25$ &$2.38$&$0$\\
		\hline
		$1.57$& $1.53$& $1.64$ &$4.91$& $5.27$& $2.38$&$0$ \\
		\hline
		$1.57$& $1.53$ &$1.64$& $4.92$& $5.25$& $2.39$&$0$ \\
		\hline
		$1.57$& $1.53$& $1.64$& $4.90$& $5.27$& $2.38$ &$0$\\
		\hline
		$1.57$ & $1.53$& $1.64$& $4.90$& $5.26$& $2.39$&$0$ \\
		\hline
		$1.57$& $1.53$ & $1.64$& $4.90$ &$5.26$& $2.40$&$0$\\
		\hline
	\end{tabular}
        \caption{Three-qubit states characterized by the angles $\theta$ and $\alpha$ are defined as tangent to the EW hyperplane.}
        \label{tab: 3 qubit alpha}
\end{table}

In this case, for the optimal condition of the EW, by performing calculations, it can be seen that there is no vector orthogonal to the kernel of EW, so that is optimal.

\subsubsection{PPTES}

To check the entangled witness of PPT states, we consider the density matrix introduced in the article \cite{Lewenstein11} which is edge state in three-qubit Hilbert space and has positive partial transpose with respect to each subsystem.

\begin{equation}
	\rho_{ppt} = \frac{1}{n} \begin{bmatrix}
		1 & 0 & 0 & 0 & 0 & 0 & 0 & 1 \\
		0 & a & 0 & 0 & 0 & 0 & 0 & 0 \\
		0 & 0 & b & 0 & 0 & 0 & 0 & 0 \\
		0 & 0 & 0 & c & 0 & 0 & 0 & 0 \\
		0 & 0 & 0 & 0 & \frac{1}{c} & 0 & 0 & 0 \\
		0 & 0 & 0 & 0 & 0 & \frac{1}{b} & 0 & 0 \\
		0 & 0 & 0 & 0 & 0 & 0 & \frac{1}{a} & 0 \\
		1 & 0 & 0 & 0 & 0 & 0 & 0 & 1 \\
	\end{bmatrix},
\end{equation}
With $n=2+a+ \frac{1}{a}+b+\frac{1}{b}+c+\frac{1}{c}$ and $a,b,c>0$. To obtain training data in this case, we choose the range of coefficients $(0<a,b,c<1)$, and in this case, using svm algorithm, we obtained the following EW for detecting these states. According to the range of coefficients of the density matrix and obtaining training data in this area, the optimal entangled witness can be obtained.

\begin{equation}\label{ew22}
	\resizebox{\textwidth}{!}{
		$\begin{aligned}
			\mathbf{W} = \begin{bmatrix}
				(0.359) & (-0.004-0.0189\textbf{i}) & (-0.0045+0.0429\textbf{i}) & (0.0358+0.008\textbf{i}) & (-0.0028-0.0009\textbf{i}) & (0.0037+0.0085\textbf{i}) & (0.0362-0.0143\textbf{i}) & (-0.4971+0.068\textbf{i}) \\
				(-0.0046+0.0189\textbf{i}) & (0.0972) & (-0.0177-0.0214\textbf{i}) & (-0.0255-0.0078\textbf{i}) & (0.0062-0.0269\textbf{i}) & (-0.0167+0.0079\textbf{i}) & (0.0248-0.0502\textbf{i}) & (-0.0148+0.0053\textbf{i}) \\
				(-0.0045-0.0429\textbf{i}) & (-0.0177+0.0214\textbf{i}) & (0.123) & (-0.015-0.0022\textbf{i}) & (-0.0043-0.0159\textbf{i}) & (0.0046-0.0226\textbf{i}) & (0.0183-0.0088\textbf{i}) & (0.026-0.0287\textbf{i}) \\
				(0.0358-0.008\textbf{i}) & (-0.0255+0.0078\textbf{i}) & (-0.015+0.0022\textbf{i}) & (0.0074) & (0.0408-0.019\textbf{i}) & (0.0263+0.017\textbf{i}) & (-0.0226+0.0633\textbf{i}) & (-0.0109-0.0015\textbf{i}) \\
				(-0.0028+0.0009\textbf{i}) & (0.0062+0.0269\textbf{i}) & (-0.0043+0.0159\textbf{i}) & (0.0408+0.019\textbf{i}) & (0.0006) & (-0.0007+0.0018\textbf{i}) & (-0.0062+0.0021\textbf{i}) & (0.0003-0.0245\textbf{i}) \\
				(0.0037-0.0085\textbf{i}) & (-0.0167-0.0079\textbf{i}) & (0.0046+0.0226\textbf{i}) & (0.0263-0.017\textbf{i}) & (-0.0007-0.0018\textbf{i}) & (0.0009) & (-0.0153-0.0065\textbf{i}) & (0.0146-0.0077\textbf{i}) \\
				(0.0362+0.0143\textbf{i}) & (0.0248+0.0502\textbf{i}) & (0.0183+0.0088\textbf{i}) & (-0.0226-0.0633\textbf{i}) & (-0.0062-0.0021\textbf{i}) & (-0.0153+0.0065\textbf{i}) & (0.0007) & (0.002+0.0038\textbf{i}) \\
				(-0.4971-0.068\textbf{i}) & (-0.0148-0.0053\textbf{i}) & (0.026+0.0287\textbf{i}) & (-0.0109+0.0015\textbf{i}) & (0.0003+0.0245\textbf{i}) & (0.0146+0.0077\textbf{i}) & (0.002-0.0038\textbf{i}) & (0.4107) \end{bmatrix}
		\end{aligned}
		$
	},
\end{equation}
The Hermitian operator \ref{ew22} has at least one negative eigenvalue and after some calculations we get
\begin{equation}\label{tr1}
\mathrm{Tr}(W \rho_{ppt}) = \frac{1}{n}(-0.2239 + \frac{0.0007}{a } + 0.0972 a + \frac{0.0009}{b } + 0.123 b + \frac{0.0006}{c } + 0.0074 c).
\end{equation}
For example if we put the parameters $a = 0.3525, b = 0.3196, c = 0.81642$ then $\mathrm{Tr}(W \rho_{ppt}) =-0.0129$.
In this case, to show that the EW is non-decomposable, it is enough that it can detect a PPT state.

\subsection{Four-qubit EW's}

The Bell basis states can be generalized for four particles with spin-$\frac{1}{2}$ and the usual four-qubit GHZ states have sixteen forms.
We consider the following Werner state to this four-qubit system

\begin{equation}\label{mat1}
\rho_{Bell(i,j,k,l)}^{W}=p |\psi_{ijkl}\rangle \langle \psi_{ijkl}| + \frac{1-p}{16} \mathbb{1}_{16} ,  0\leq p \leq 1,
\end{equation}
and $|\psi_{0000}\rangle= \frac{1}{\sqrt{2}}(|0000\rangle + |1111\rangle)$ which is maximum entangled pure Bell state and in the convex set of entangled quantum states as extremal points, in this case, using the Werner density matrix, we construct training data and obtain the EW to detect these states.

\begin{equation}
\mathit{W} = \begin{bmatrix}
A & B \\
C & D
\end{bmatrix} \label{w4}
\end{equation}

\begin{equation*} 
	\resizebox{\textwidth}{!}{
		$\begin{aligned}
			\mathbf{A} = \begin{bmatrix}
				(0.059) & (0.002+0.001\textbf{i}) & (0.001+0.003\textbf{i}) & -0.007\textbf{i} & (-0-0.002\textbf{i}) & (-0.003) & 0\textbf{i} & (0.001+0.006\textbf{i}) \\
				(0.002-0.001\textbf{i}) & (0.065) & (0.113) & (0.003-0.003\textbf{i}) & (0.005+0.006\textbf{i}) & (0.001+0.002\textbf{i}) & (-0.004-0.002\textbf{i}) & (-0.004+0.009\textbf{i})  \\
				(0.001-0.003\textbf{i}) & (0.113) & (0.063) & (-0.001-0.004\textbf{i}) & (-0.001-0.003\textbf{i}) & (-0.004-0.002\textbf{i}) & (0.001-0.001\textbf{i}) & (0.01-0.004\textbf{i}) \\
				0.007\textbf{i} & (0.003+0.003\textbf{i}) & (-0.001+0.004\textbf{i}) & (0.056) & (0.001-0.004\textbf{i}) & (0.004-0.006\textbf{i}) & (0.004-0.005\textbf{i}) & (0.001) \\
				0.002\textbf{i} & (0.005-0.006\textbf{i}) & (-0.001+0.003\textbf{i}) & (0.001+0.004\textbf{i}) & (0.065) & (-0.003) & (-0.001+0.001\textbf{i}) & (-0.002+0.007\textbf{i})  \\
				(-0.003) & (0.001-0.002\textbf{i}) & (-0.004+0.002\textbf{i}) & (0.004+0.006\textbf{i}) & (-0.003) & (0.061) & (0.11) & (-0.001-0.01\textbf{i}) \\
				0j & (-0.004+0.002\textbf{i}) & (0.001+0.001\textbf{i}) & (0.004+0.005\textbf{i}) & (-0.001-0.001\textbf{i}) & (0.11) & (0.058) & (-0.009+0.004\textbf{i}) \\
				(0.001-0.006\textbf{i}) & (-0.004-0.009\textbf{i}) & (0.01+0.004\textbf{i}) & (0.001) & (-0.002-0.007\textbf{i}) & (-0.001+0.01\textbf{i}) & (-0.009-0.004\textbf{i}) & (0.072) \\
				 \end{bmatrix}
		\end{aligned}
		$
	},
\end{equation*}
\begin{equation*}
	\resizebox{\textwidth}{!}{
		$\begin{aligned}
			\mathbf{B} = \begin{bmatrix}
				 (-0-0.001\textbf{i}) & (0.011+0.005\textbf{i}) & (-0.001) & 0.001\textbf{i} & (0.004-0.002\textbf{i}) & (0.012-0.001\textbf{i}) & (0.001+0.005\textbf{i}) & (-0.607+0.002\textbf{i}) \\
				 (0.014-0.003\textbf{i}) & (0.002+0.003\textbf{i}) & (0.002+0.005\textbf{i}) & (-0-0.001\textbf{i}) & (0.001+0.001\textbf{i}) & (0.002-0.002\textbf{i}) & (-0-0.006\textbf{i}) & (-0.005+0.005\textbf{i}) \\
				 (-0.001) & (0.002+0.005\textbf{i}) & (0.003) & (-0.001+0.006\textbf{i}) & (-0.005) & (-0-0.006\textbf{i}) & (0.001-0.002\textbf{i}) & (-0.007-0.008\textbf{i}) \\
				 (-0.01+0.001\textbf{i}) & (-0.005+0.006\textbf{i}) & (-0.001+0.005\textbf{i}) & (0.001+0.001\textbf{i}) & (0.004-0.01\textbf{i}) & (0.003-0.004\textbf{i}) & (-0.005-0.009\textbf{i}) & 0.004\textbf{i} \\
			    (0.002-0.003\textbf{i}) & (0.005+0.005\textbf{i}) & (0.007-0.001\textbf{i}) & (0.008+0.005\textbf{i}) & (0.001-0.001\textbf{i}) & (0.003) & (0.005+0.002\textbf{i}) & (0.005+0.001\textbf{i}) \\
				 (-0.007-0.005\textbf{i}) & (-0.006-0.008\textbf{i}) & (0.002-0.021\textbf{i}) & (-0.005+0.01\textbf{i}) & (-0.002+0.001\textbf{i}) & (-0.003-0.001\textbf{i}) & (-0.002-0.003\textbf{i}) & (-0.007-0.003\textbf{i}) \\
				 (-0.004+0.003\textbf{i}) & (0.002-0.021\textbf{i}) & (0.004-0.001\textbf{i}) & (0.018+0.012\textbf{i}) & (-0.001-0.001\textbf{i}) & (-0.002-0.003\textbf{i}) & (0.001-0.003\textbf{i}) & (-0.007-0.003\textbf{i}) \\
				 0.002\textbf{i} & (-0-0.008\textbf{i}) & (-0.008+0.006\textbf{i}) & (0.011-0.003\textbf{i}) & (-0.011-0.001\textbf{i}) & (-0.007-0.001\textbf{i}) & (0.009-0.007\textbf{i}) & (0.002) \\
 \end{bmatrix}
		\end{aligned}
		$
	},
\end{equation*}
\begin{equation*}
	\resizebox{\textwidth}{!}{
		$\begin{aligned}
			\mathbf{C} = \begin{bmatrix}
			
				0.001\textbf{i} & (0.014+0.003\textbf{i}) & (-0.001) & (-0.01-0.001\textbf{i}) & (0.002+0.003\textbf{i}) & (-0.007+0.005\textbf{i}) & (-0.004-0.003\textbf{i}) & -0.002\textbf{i}   \\
				(0.011-0.005\textbf{i}) & (0.002-0.003\textbf{i}) & (0.002-0.005\textbf{i}) & (-0.005-0.006\textbf{i}) & (0.005-0.005\textbf{i}) & (-0.006+0.008\textbf{i}) & (0.002+0.021\textbf{i}) & 0.008\textbf{i}  \\
				(-0.001) & (0.002-0.005\textbf{i}) & (0.003) & (-0.001-0.005\textbf{i}) & (0.007+0.001\textbf{i}) & (0.002+0.021\textbf{i}) & (0.004+0.001\textbf{i}) & (-0.008-0.006\textbf{i})  \\
				-0.001\textbf{i} & 0.001\textbf{i} & (-0.001-0.006\textbf{i}) & (0.001-0.001\textbf{i}) & (0.008-0.005\textbf{i}) & (-0.005-0.01\textbf{i}) & (0.018-0.012\textbf{i}) & (0.011+0.003\textbf{i})  \\
				(0.004+0.002\textbf{i}) & (0.001-0.001\textbf{i}) & (-0.005) & (0.004+0.01\textbf{i}) & (0.001+0.001\textbf{i}) & (-0.002-0.001\textbf{i}) & (-0.001+0.001\textbf{i}) & (-0.011+0.001\textbf{i})  \\
				(0.012+0.001\textbf{i}) & (0.002+0.002\textbf{i}) & 0.006\textbf{i} & (0.003+0.004\textbf{i}) & (0.003) & (-0.003+0.001\textbf{i}) & (-0.002+0.003\textbf{i}) & (-0.007+0.001\textbf{i}) \\
				(0.001-0.005\textbf{i}) & 0.006\textbf{i} & (0.001+0.002\textbf{i}) & (-0.005+0.009\textbf{i}) & (0.005-0.002\textbf{i}) & (-0.002+0.003\textbf{i}) & (0.001+0.003\textbf{i}) & (0.009+0.007\textbf{i})  \\
				(-0.607-0.002\textbf{i}) & (-0.005-0.005\textbf{i}) & (-0.007+0.008\textbf{i}) & (-0-0.004\textbf{i}) & (0.005-0.001\textbf{i}) & (-0.007+0.003\textbf{i}) & (-0.007+0.003\textbf{i}) & (0.002)
				 \end{bmatrix}
		\end{aligned}
		$
	},
\end{equation*}

\begin{equation*}
	\resizebox{\textwidth}{!}{
		$\begin{aligned}
			\mathbf{D} = \begin{bmatrix}

				 (0.065) & (0.002) & (0.005-0.003\textbf{i}) & (0.003-0.006\textbf{i}) & (0.003+0.003\textbf{i}) & (-0.004-0.004\textbf{i}) & (-0.001-0.006\textbf{i}) & (0.001-0.001\textbf{i}) \\
				 (0.002) & (0.061) & (0.101) & 0.002\textbf{i} & (-0.006) & (-0.002-0.001\textbf{i}) & (-0.004+0.005\textbf{i}) & (-0-0.005\textbf{i}) \\
				 (0.005+0.003\textbf{i}) & (0.101) & (0.061) & (0.003+0.009\textbf{i}) & (-0.008-0.01\textbf{i}) & (-0.004+0.005\textbf{i}) & (-0.001) & (-0.007-0.005\textbf{i}) \\
				 (0.003+0.006\textbf{i}) & -0.002\textbf{i} & (0.003-0.009\textbf{i}) & (0.069) & (0.003+0.005\textbf{i}) & (-0.003+0.007\textbf{i}) & (-0.003+0.003\textbf{i}) & (0.004+0.003\textbf{i}) \\
				 (0.003-0.003\textbf{i}) & (-0.006) & (-0.008+0.01\textbf{i}) & (0.003-0.005\textbf{i}) & (0.062) & (-0.002) & (-0.004+0.001j\textbf{i}) & (-0.007-0.005\textbf{i}) \\
				 (-0.004+0.004\textbf{i}) & (-0.002+0.001\textbf{i}) & (-0.004-0.005\textbf{i}) & (-0.003-0.007\textbf{i}) & (-0.002) & (0.063) & (0.104) & 0\textbf{i} \\
				 (-0.001+0.006\textbf{i}) & (-0.004-0.005\textbf{i}) & (-0.001) & (-0.003-0.003\textbf{i}) & (-0.004-0.001\textbf{i}) & (0.104) & (0.067) & 0.005\textbf{i} \\
				 (0.001+0.001\textbf{i}) & 0.005\textbf{i} & (-0.007+0.005\textbf{i}) & (0.004-0.003\textbf{i}) & (-0.007+0.005\textbf{i}) & 0\textbf{i} & -0.005\textbf{i} & (0.055)
				     \end{bmatrix}
		\end{aligned}
		$
	},
\end{equation*}

\begin{equation}
	\text{Tr}(W \rho_{GHZ(i,j,k,l)}^{W})=-0.612 p+0.062.
\end{equation}
In the interval  $ 0.11\leq p \leq 1$, the values of $p$ represent the entangled region where Tr$(W \rho_{GHZ(i,j,k,l)}^{W})<0$  and  $ p < 0.11 $, defined as separable region with Tr$(W \rho_{GHZ(i,j,k,l)}^{W})>0$.

\begin{figure}[H]
	\centering
 	\includegraphics[width=0.7\linewidth]{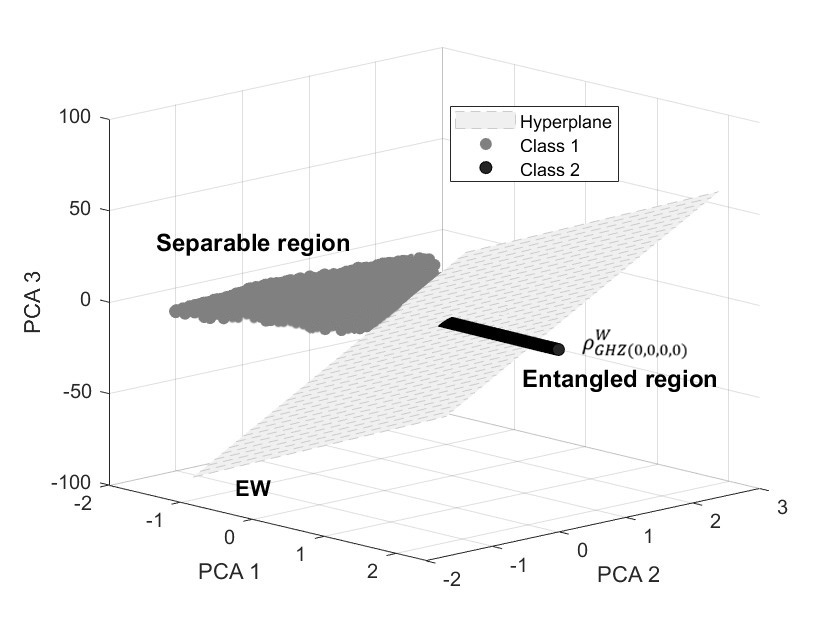}
	\caption{Principal component analysis (PCA) plots of three principal components of separable and entangled states of four-qubit states. PC1, principal 1, PC2, principal 2, PC3, Principal 3 and optimal entangled witness for a four-qubit density matrice ${\rho_{GHZ(0,0,0,0)}^{W}}$.}
	
\end{figure}

\begin{table}[H]
	\centering
	\begin{tabular}{|c|c|c|c|c|c|c|c|c|}
		\hline
		$\theta_1$ & $\theta_2$ &$\theta_3$ &$\theta_4$ & $\alpha_1$ & $\alpha_2$ &$\alpha_3$ &$\alpha_4$& Tr$(W_{full ent} |\nu\rangle \langle \nu |)$ \\
		\hline
		$2.51$& $2.51$& $1.88$& $1.25$& $4.71$& $3.14$& $4.71$& $1.57$&$0$ \\
		\hline
		$1.25$&$1.25$&$1.88$& $1.25$& $3.14$&$ 1.57$&$3.14$&$4.71$&$0$ \\
		\hline
		$0.62$&$ 2.51$& $1.88$& $1.88$&$4.71$&$ 1.57$&$ 1.57$& $ 4.71$&$0$\\
		\hline
		$0.62$& $0.62$& $1.25$& $1.88$& $4.71$&$ 3.14$&$4.71$&$1.57$&$0$ \\
		\hline
		$0$& $1.25$& $1.88$& $1.88$& $0$&$3.14$&$ 4.71$& $1.57$&$0$ \\
		\hline
		$ 0$&$1.25$& $1.25$& $1.88$&$ 0$&$3.14$&$ 4.71$& $1.57$ &$0$\\
		\hline
		$1.61$&$1.61$& $1.61$& $1.61$&$4.21$&$ 4.62$&$ 2.70$&$5.92$&$0$ \\
		\hline
		$1.61$& $1.61$& $1.61$&$1.61$& $ 4.21$&$ 4.03$&$ 4.62$&$5.92$&$0$\\
		\hline
		$1.61$&$1.61$&$1.61$& $1.61$&$ 4.03$&$ 4.03$&$4.62$&$ 5.92$&$0$ \\
		\hline
		$1.61$& $ 1.61$&$1.61$& $1.61$&$5.92$&$ 4.62$&$4.62$&$2.70$&$0$ \\
		\hline
		$1.61$&$1.61$& $1.61$& $ 1.61$& $5.92$&$ 2.70$&$4.62$& $5.92$&$0$ \\
		\hline
		$1.61$& $1.61$& $1.61$&$1.78$&$4.62$&$ 4.21$&$2.70$&$5.92$&$0$ \\
		\hline
		$1.61$&$1.61$&$1.61$& $1.78$& $4.21$&$ 4.62$&$2.70$& $5.92$&$0$\\
		\hline
		$1.61$&$1.61$&$1.78$& $1.61$& $2.70$&$ 5.92$&$5.92$&$4.62$&$0$ \\
		\hline
		$1.61$&$1.78$& $1.78$&$1.67$& $4.03$&$ 4.03$&$5.92$&$4.21$&$0$ \\
		\hline
		$2.73$&$ 2.73$&$1.81$&$1.81$&$5.92$&$4.21$&$5.92$&$2.70$&$0$ \\
		\hline	
	\end{tabular}
        \caption{These four-qubit states, defined by the angles $\theta$ and $\alpha$, are identified as being tangent to the EW hyperplane.}
\end{table}
In another example for the case where
\begin{equation}\label{mat321}
\rho_{GHZ(i,j,k,l)}^{W}=p |\psi_{ijkl}\rangle \langle \psi_{ijkl} | + \frac{1-p}{16} \mathbb{1}_{16} ,  0\leq p \leq 1,
\end{equation}
where
\begin{equation} |\psi \rangle = \frac{1}{\sqrt{2}} (|000 \rangle+|111 \rangle) \otimes (\cos\theta |0 \rangle +e^{i\alpha}\sin\theta |1 \rangle),
\end{equation}
and
\begin{equation}
	\text{Tr}(W\rho_{GHZ(i,j,k,l)}^{W}) =  - 1.7 \times 10^{-1} p + 6.2 \times 10^{-2}.
\end{equation}

We consider the parameter range $0.35\leq p \leq 1$ for the creation of entangled states in a bi-partite system. This parameter regime is identified by analyzing the real part of a specific expression within the intervals $0\leq\alpha\leq 2\pi$ and $0\leq\theta\leq \pi$. The key indicator of entanglement in this context is the trace negativity of a specific operator W acting on a four-qubit bi-partite state.

\begin{equation*}
	\resizebox{\textwidth}{!}{
		$\begin{aligned}
			\mathbf{A} = \begin{bmatrix}
				(0.04) & (0.001+0.001\textbf{i}) & (-0.002+0.003\textbf{i}) & (0.003-0.002\textbf{i}) & (-0.001-0.002\textbf{i}) & -0.003\textbf{i} & 0.001\textbf{i} & (-0-0.001\textbf{i})  \\
				(0.001-0.001\textbf{i}) & (0.04) & (0.018) & (-0.003+0.001\textbf{i}) & (0.001-0.003\textbf{i}) & (-0.004+0.002\textbf{i}) & (-0-0.001\textbf{i}) & (0.005-0.007\textbf{i})  \\
				(-0.002-0.003\textbf{i}) & (0.018) & (0.07) & -0.003\textbf{i} & (-0-0.002\textbf{i}) & (-0-0.001\textbf{i}) & (0.002-0.003\textbf{i}) & 0\textbf{i} \\
				(0.003+0.002\textbf{i}) & (-0.003-0.001\textbf{i}) & 0.003\textbf{i} & (0.076) & (0.001-0.003\textbf{i}) & (-0.001+0.002\textbf{i}) & (-0.001-0.002\textbf{i}) & (0.001+0.002\textbf{i})  \\
				(-0.001+0.002\textbf{i}) & (0.001+0.003\textbf{i}) & 0.002\textbf{i} & (0.001+0.003\textbf{i}) & (0.061) & (0.003) & (0.002+0.004\textbf{i}) & (0.001)  \\
				0.003\textbf{i} & (-0.004-0.002\textbf{i}) & 0.001\textbf{i} & (-0.001-0.002j\textbf{i}) & (0.003) & (0.079) & (0.011) & (0.002-0.002\textbf{i}) \\
				-0.001\textbf{i} & 0.001\textbf{i} & (0.002+0.003\textbf{i}) & (-0.001+0.002\textbf{i}) & (0.002-0.004\textbf{i}) & (0.011) & (0.066) & (0.001+0.001\textbf{i})  \\
				0.001\textbf{i} & (0.005+0.007\textbf{i}) & (-0) & (0.001-0.002\textbf{i}) & (0.001) & (0.002+0.002\textbf{i}) & (0.001-0.001\textbf{i}) & (0.069) \\
				 \end{bmatrix}
		\end{aligned}
		$
	},
\end{equation*}

\begin{equation*}
	\resizebox{\textwidth}{!}{
		$\begin{aligned}
			\mathbf{B}= \begin{bmatrix}
				 (-0.001-0.002\textbf{i}) & (0.002-0.001\textbf{i}) & (0.002-0.002\textbf{i}) & (0.001+0.003\textbf{i}) & (-0.001-0.003\textbf{i}) & (0.012+0.002\textbf{i}) & (-0.152+0.001\textbf{i}) & (-0.001+0.001\textbf{i}) \\
				 0.001\textbf{i} & (0.002-0.003\textbf{i}) & (0.001-0.001\textbf{i}) & (0.001+0.003\textbf{i}) & (-0.001) & (-0.002-0.002\textbf{i}) & (-0.001-0.001\textbf{i}) & (-0.153+0.002\textbf{i}) \\
				 (-0.005+0.002\textbf{i}) & (0.001-0.001\textbf{i}) & (-0.001) & 0.001\textbf{i} & (-0.002-0.008\textbf{i}) & 0.001\textbf{i} & (-0.002+0.002\textbf{i}) & (0.011+0.001\textbf{i}) \\
				 0 & (-0.004+0.001\textbf{i}) & 0.002\textbf{i} & (0.003-0.001\textbf{i}) & (0.002-0.002\textbf{i}) & (0.001+0.007\textbf{i}) & (-0-0.001\textbf{i}) & 0.001\textbf{i} \\
				(-0.007-0.003\textbf{i}) & (0.001+0.002\textbf{i}) & (0.002+0.001\textbf{i}) & (0.002) & (-0.001+0.004\textbf{i}) & 0\textbf{i} & 0.003\textbf{i} & (-0.001+0.003\textbf{i}) \\
			    (0.002) & (0.002+0.003\textbf{i}) & (-0) & (0.004+0.005\textbf{i}) & 0.001\textbf{i} & (0.001) & (-0.001-0.001\textbf{i}) & (-0.002-0.001\textbf{i}) \\
				 (0.003-0.001\textbf{i}) & (-0) & (-0.009-0.006\textbf{i}) & (0.002+0.001\textbf{i}) & (0.005+0.006\textbf{i}) & (-0.001-0.001\textbf{i}) & (-0.005+0.005\textbf{i}) & 0.003\textbf{i} \\
				 (0.002) & 0.004\textbf{i} & (0.003) & (-0.001-0.003\textbf{i}) & (-0.001+0.001\textbf{i}) & (-0.001-0.001\textbf{i}) & -0.002\textbf{i} & (0.004-0.002\textbf{i}) \\
				 \end{bmatrix}
		\end{aligned}
		$
	},
\end{equation*}

\begin{equation*}
	\resizebox{\textwidth}{!}{
		$\begin{aligned}
			\mathbf{C} = \begin{bmatrix}
				(-0.001+0.002\textbf{i}) & (-0-0.001\textbf{i}) & (-0.005-0.002\textbf{i}) & 0\textbf{i} & (-0.007+0.003\textbf{i}) & (0.002) & (0.003+0.001\textbf{i}) & (0.002)  \\
				(0.002+0.001\textbf{i}) & (0.002+0.003\textbf{i}) & (0.001+0.001\textbf{i}) & (-0.004-0.001\textbf{i}) & (0.001-0.002\textbf{i}) & (0.002-0.003\textbf{i}) & 0\textbf{i} & (-0-0.004\textbf{i}) \\
				(0.002+0.002\textbf{i}) & (0.001+0.001\textbf{i}) & (-0.001) & (-0-0.002\textbf{i}) & (0.002-0.001\textbf{i}) & 0\textbf{i} & (-0.009+0.006\textbf{i}) & (0.003) \\
				(0.001-0.003\textbf{i}) & (0.001-0.003\textbf{i}) & (-0-0.001\textbf{i}) & (0.003+0.001\textbf{i}) & (0.002) & (0.004-0.005\textbf{i}) & (0.002-0.001\textbf{i}) & (-0.001+0.003\textbf{i})  \\
				(-0.001+0.003\textbf{i}) & (-0.001) & (-0.002+0.008\textbf{i}) & (0.002+0.002\textbf{i}) & (-0.001-0.004\textbf{i}) & -0.001\textbf{i} & (0.005-0.006\textbf{i}) & (-0.001-0.001\textbf{i})  \\
				(0.012-0.002\textbf{i}) & (-0.002+0.002\textbf{i}) & -0.001\textbf{i}& (0.001-0.007\textbf{i}) & (-0) & (0.001) & (-0.001+0.001\textbf{i}) & (-0.001+0.001\textbf{i}) \\
				(-0.152-0.001\textbf{i}) & (-0.001+0.001\textbf{i}) & (-0.002-0.002\textbf{i}) & 0.001\textbf{i} & (-0-0.003\textbf{i}) & (-0.001+0.001\textbf{i}) & (-0.005-0.005\textbf{i}) & 0.002\textbf{i}  \\
				(-0.001-0.001\textbf{i}) & (-0.153-0.002\textbf{i}) & (0.011-0.001\textbf{i}) & (-0-0.001\textbf{i}) & (-0.001-0.003\textbf{i}) & (-0.002+0.001\textbf{i}) & (-0-0.003\textbf{i}) & (0.004+0.002\textbf{i})
				 \end{bmatrix}
		\end{aligned}
		$
	},
\end{equation*}

\begin{equation*}
	\resizebox{\textwidth}{!}{
		$\begin{aligned}
			\mathbf{D} = \begin{bmatrix}
				 (0.076) & -0.002\textbf{i} & (-0.003-0.004\textbf{i}) & 0.001\textbf{i} & (-0.003-0.001\textbf{i}) & 0.001\textbf{i} & (-0.001) & -0.002\textbf{i} \\
				 0.002\textbf{i}& (0.072) & (0.013) & (-0.003+0.004\textbf{i}) & (0.002-0.002\textbf{i}) & (0.002-0.001\textbf{i}) & 0\textbf{i} & (-0.002+0.002\textbf{i}) \\
				 (-0.003+0.004\textbf{i}) & (0.013) & (0.072) & (-0.001) & (-0.002-0.002\textbf{i}) & 0\textbf{i} & (0.002) & (0.003+0.003\textbf{i}) \\
				 (-0-0.001\textbf{i}) & (-0.003-0.004\textbf{i}) & (-0.001) & (0.068) & (0.001-0.001\textbf{i}) & (-0.003-0.001\textbf{i}) & (-0.003) & (-0.004+0.001\textbf{i}) \\
				 (-0.003+0.001\textbf{i}) & (0.002+0.002\textbf{i}) & (-0.002+0.002\textbf{i}) & (0.001+0.001\textbf{i}) & (0.063) & (0.001) & 0.001\textbf{i} & (-0.006-0.003\textbf{i}) \\
				 (-0-0.001\textbf{i}) & (0.002+0.001\textbf{i}) & (-0) & (-0.003+0.001\textbf{i}) & (0.001) & (0.072) & (0.022) & (-0.001+0.003\textbf{i}) \\
				 (-0.001) & (-0) & (0.002) & (-0.003) & (-0-0.001\textbf{i}) & (0.022) & (0.04) & (-0-0.002\textbf{i}) \\
				 0.002\textbf{i} & (-0.002-0.002\textbf{i}) & (0.003-0.003\textbf{i}) & (-0.004-0.001\textbf{i}) & (-0.006+0.003\textbf{i}) & (-0.001-0.003\textbf{i}) & 0.002\textbf{i} & (0.037) \end{bmatrix}
		\end{aligned}
		$
	}.
\end{equation*}

\section{Conclusions}
In this paper, we have obtained hyperplanes tangent to the region of separable states for the two, three and four-qubit systems and can be generalized to N-qubit, which are the same as EW, using the machine learning algorithm. We have proved that all the EW's are completely tangent to the separable region and are therefore optimal. Also, in the three-qubit space, we have obtained non-decomposable EW's to detect PPT entangled states.

\section{Appendix}

\begin{algorithm}
\caption{Entanglement Witness Using Primal Support Vector Machine }
\begin{algorithmic}

\STATE \textbf{Inputs:} 
\begin{itemize}
    \item Regularization parameter $C$.
    \item Training dataset $\{(x_i, y_i)\}$, where $x_i \in \mathbb{R}^n$, $y_i \in \{-1, 1\}$.
\end{itemize}

\STATE \textbf{Step 1: Data Preparation and Measurement}
\begin{itemize}
    \item Construct density matrices representing quantum states.
    \item Perform quantum measurements using Pauli matrices to map the quantum state features into a classical feature space.
\end{itemize}

\STATE \textbf{Step 2: Define the Primal Optimization Problem}
\[
\min_{w, b, \xi} \quad \frac{1}{2} \|w\|^2 + C \sum_{i=1}^N \xi_i,
\]
\[
\text{subject to: } \quad y_i (w^\top x_i + b) \geq 1 - \xi_i, \quad \xi_i \geq 0, \quad \forall i \in \{1, \dots, N\}.
\]

\STATE \textbf{Step 3: Solve the Optimization Problem}
\begin{itemize}
    \item Initialize $w$, $b$, and $\xi_i$ (commonly set to zero).
    \item Optimize the primal objective function $L$ using gradient descent or quadratic programming:
    \begin{itemize}
        \item Update $w$: $w \leftarrow w - \eta \frac{\partial L}{\partial w}$.
        \item Update $b$: $b \leftarrow b - \eta \frac{\partial L}{\partial b}$.
        \item Update $\xi_i$ as necessary to satisfy constraints.
    \end{itemize}
    \item Classify a test sample $x$ using:
    \[
    \hat{y} = \text{sign}(w^\top x + b),
    \]
    where $w$ defines the hyperplane.
    \item Evaluate performance on a test dataset and fine-tune $C$ to optimize generalization.
\end{itemize}

\end{algorithmic}
\end{algorithm}

\clearpage
\begin{algorithm}
\begin{algorithmic}
\STATE \textbf{Step 4: Establish the Entanglement Witness Operator}
\begin{itemize}
    \item Translate the SVM decision boundary into an entanglement witness.
    \item For a two-qubit system, define:
    \[
    W = \sum_{i_1, i_2} c_{i_1, i_2} \sigma_{i_1} \otimes \sigma_{i_2},
    \]
    where $\sigma_{i_j}$ are the Pauli matrices $(\sigma_x, \sigma_y, \sigma_z)$ and the identity matrix $I$, and $c_{i_1, i_2}$ are hyperplane's coefficients derived from the SVM solution.
\end{itemize}

\STATE \textbf{Output:}
The SVM decision boundary $(w, b)$ is used to define the entanglement witness $W$, ensuring it satisfies the necessary criteria for detecting quantum entanglement.
\end{algorithmic}
\end{algorithm}

\bibliographystyle{ieeetr}
\bibliography{refMLL}

\begin{thebibliography}{10}

\bibitem{Einstein}
B.~P. A.~Einstein and N.~Rosen, ``Can quantum-mechanical description of
  physical reality be considered complete?,'' {\em Phys. Rev.}, vol.~47,
  no.~777, 1935.

\bibitem{Schrodinger}
E.~Schrodinger, ``Discussion of probability relations between separated
  systems,'' {\em Proc. Cambridge Philos. Soc.}, vol.~31, no.~555, 1935.

\bibitem{Nielsen}
M.~A. Nielsen and I.~L. Chuang, {\em Quantum Computation and Quantum
  Information}.
\newblock Cambridge University Press, 2000.

\bibitem{Bennett}
C.~H. Bennett, G.~Brassard, C.~Cr{\'e}peau, R.~Jozsa, A.~Peres, and W.~K.
  Wootters, ``Teleporting an unknown quantum state via dual classical and
  einstein-podolsky-rosen channels,'' {\em Phys. Rev. Lett}, vol.~70, p.~1895,
  1993.

\bibitem{Bennett1}
C.~H. Bennett and S.~J. Wiesner, ``Communication via oneand two-particle
  operators on einstein-podolsky-rosen states,'' {\em Phys. Rev. Lett},
  vol.~69, p.~2881, 1992.

\bibitem{PhysRevLett.67.661}
A.~K. Ekert, ``Quantum cryptography based on bell's theorem,'' {\em Phys. Rev.
  Lett}, vol.~67, pp.~661--663, Aug 1991.

\bibitem{GUHNE20091}
O.~Gühne and G.~Tóth, ``Entanglement detection,'' {\em Physics Reports},
  vol.~474, no.~1, pp.~1--75, 2009.

\bibitem{Gurvits}
L.~Gurvits, ``Classical deterministic complexity of edmonds’ problem and
  quantum entanglement,'' {\em In Proceedings of the Thirty-Fifth Annual ACM
  Symposium on Theory of Computing, STOC}, vol.~03, no.~10-19, 2003.

\bibitem{Horodecki22}
M.~H. R.~Horodecki, P.~Horodecki and K.~Horodecki, ``Quantum entanglement,''
  {\em Rev. Mod. Phys.}, vol.~81, p.~865, 2009.

\bibitem{Gabriel}
B.~C.~H. A.~Gabriel and M.~Huber, ``Criterion for k separability in mixed
  multipartite states,'' {\em Quantum Inf. Comput.}, vol.~10, p.~0829, 2010.

\bibitem{Peres}
A.~Peres, ``Separability criterion for density matrices,'' {\em Phys. Rev.
  Lett.}, vol.~77, p.~1413, 1996.

\bibitem{Horodecki}
M.~Horodecki, P.~Horodecki, and R.~Horodecki, ``Separability of mixed states:
  necessary and sufficient conditions,'' {\em Phys. Lett. A}, vol.~223, no.~2,
  p.~1, 1996.

\bibitem{Horodecki12}
P.~Horodecki, ``Separability criterion and inseparable mixed states with
  positive partial transposition,'' {\em Phys. Lett. A}, vol.~232, no.~5,
  p.~333, 1997.

\bibitem{Lewenstein}
M.~Lewenstein, B.~Kraus, J.~I. Cirac, and P.~Horodecki, ``Optimization of
  entanglement witnesses,'' {\em Phys. Rev. A}, vol.~62, p.~052310, 2000.

\bibitem{Jafarizadeh}
A.~H. M.A.~Jafarizadeh, M.~Mahdian and K.~Aghayar, ``Detecting some three-qubit
  mub diagonal entangled states via nonlinear optimal entanglement witnesses,''
  {\em Eur. Phys. J. D}, vol.~50, p.~107, 2008.

\bibitem{PhysRevA.78.032313}
M.~A. Jafarizadeh, Y.~Akbari, K.~Aghayar, A.~Heshmati, and M.~Mahdian,
  ``Investigating a class of
  $2\ensuremath{\bigotimes}2\ensuremath{\bigotimes}d$ bound entangled density
  matrices via linear and nonlinear entanglement witnesses constructed by exact
  convex optimization,'' {\em Phys. Rev. A}, vol.~78, p.~032313, Sep 2008.

\bibitem{lewenstein2000optimization}
M.~Lewenstein, B.~Kraus, J.~I. Cirac, and P.~Horodecki, ``Optimization of
  entanglement witnesses,'' {\em Physical Review A}, vol.~62, no.~5, p.~052310,
  2000.

\bibitem{Arunachalam}
R.~G.~c. Arunachalam, S. de~Wolf, ``a survey of quantum learning theory.,''
  {\em SIGACT News}, vol.~48, no.~41-67, 2017.

\bibitem{Ciliberto}
C.~e.~a. Ciliberto, ``Quantum machine learning,'' {\em a classical perspective.
  Proc. R. Soc. Lond. A}, vol.~474, no.~20170551, 2018.

\bibitem{Chrisley}
R.~Chrisley, ``Quantum learning, in new directions in cognitive science:
  Proceedings of the international symposium, saariselka, finland, p. pylkknen
  and p. pylkk (editors),'' no.~77-89, 1995.

\bibitem{Siheon}
e.~a. Siheon~Park, ``Variational quantum approximate support vector machine
  with inference transfer,'' {\em Scientific Reports.}, vol.~13, no.~3288,
  2023.

\bibitem{Torlai}
G.~e.~a. Torlai, ``Neural-network quantum state tomography,'' {\em Nat. Phys.},
  vol.~14, no.~447–450, 2018.

\bibitem{PhysRevX.8.031086}
M.~Bukov, A.~G.~R. Day, D.~Sels, P.~Weinberg, A.~Polkovnikov, and P.~Mehta,
  ``Reinforcement learning in different phases of quantum control,'' {\em Phys.
  Rev. X}, vol.~8, p.~031086, Sep 2018.

\bibitem{Ming}
Y.~Ming, C.-T. Lin, S.~D. Bartlett, and W.-W. Zhang, ``Quantum topology
  identification with deep neural networks and quantum walks,'' {\em npj
  Comput. Mater}, vol.~5, 2019.

\bibitem{Carleo}
G.~Carleo and M.~Troyer, ``Solving the quantum many-body problem with
  artificial neural networks,'' {\em Science,}, vol.~355, no.~602–606, 2017.

\bibitem{Naema}
e.~a. Naema~Asif, ``Entanglement detection with artificial neural networks,''
  {\em Scientific Reports.}, vol.~13, no.~1562, 2023.

\bibitem{PhysRevResearch.3.033135}
J.~Y. Khoo and M.~Heyl, ``Quantum entanglement recognition,'' {\em Phys. Rev.
  Res.}, vol.~3, p.~033135, Aug 2021.

\bibitem{Harney_2021}
C.~Harney, M.~Paternostro, and S.~Pirandola, ``Mixed state entanglement
  classification using artificial neural networks,'' {\em New Journal of
  Physics}, vol.~23, p.~063033, jun 2021.

\bibitem{PhysRevA.98.012315}
S.~Lu, S.~Huang, K.~Li, J.~Li, J.~Chen, D.~Lu, Z.~Ji, Y.~Shen, D.~Zhou, and
  B.~Zeng, ``Separability-entanglement classifier via machine learning,'' {\em
  Phys. Rev. A}, vol.~98, p.~012315, Jul 2018.

\bibitem{8764462}
P.-H. Qiu, X.-G. Chen, and Y.-W. Shi, ``Detecting entanglement with deep
  quantum neural networks,'' {\em IEEE Access}, vol.~7, pp.~94310--94320, 2019.

\bibitem{support}
D.~A. Pisner and D.~M. Schnyer, ``Support vector machine,'' in {\em Machine
  learning}, pp.~101--121, Elsevier, 2020.

\bibitem{svm}
N.~Cristianini and J.~Shawe-Taylor, {\em An introduction to support vector
  machines and other kernel-based learning methods}.
\newblock Cambridge university press, 2000.

\bibitem{Giuntini}
R.~Giuntini, H.~Freytes, D.~K. Park, C.~Blank, F.~Holik, K.~L. Chow, and
  G.~Sergioli, ``Quantum state discrimination for supervised classification,''
  {\em https://arxiv.org/abs/2104.00971}, 2021.

\bibitem{Lewenstein11}
A.~Ac{\'\i}n, D.~Bru{\ss}, M.~Lewenstein, and A.~Sanpera, ``Classification of
  mixed three-qubit states,'' {\em Phys. Rev. Lett.}, vol.~87, no.~4, 2001.

\bibitem{Vidal}
W.~D{\"u}r, G.~Vidal, and J.~I. Cirac, ``Three qubits can be entangled in two
  inequivalent ways,'' {\em Phys. Rev. A.}, vol.~62, no.~062314, 2000.

\end{thebibliography}
\end{document}